 \documentclass[aps,prl,twocolumn,groupedaddress,citeautoscript,nopacs]{revtex4-1}
\usepackage{graphicx}
\usepackage{latexsym}
\usepackage{amsmath}
\usepackage{amsthm}
\usepackage{amssymb}
\usepackage{epstopdf} 
\usepackage{enumerate}
\usepackage{setspace} 
\usepackage{dcolumn}
\usepackage{bm}
\usepackage{setspace} 
\usepackage{slashed}
\usepackage{color}
\usepackage{youngtab}

\begin{document}
\title{  Deconfined Thermal Phase Transitions \\ with $Z_2$ Gauge Structures
}

 \author{Eun-Gook Moon }
\affiliation{Department of Physics, Korea Advanced Institute of Science and Technology, Daejeon 305-701, Korea}

\date{\today}

\begin{abstract}  
Fathoming deconfined phases is one of the key issues in modern condensed matter. Striking many-body effects including massive quantum entanglement and coherence may be realized as manifested in quantum spin liquids and topological orders. Here, we demonstrate that deconfined phases even host exotic thermal phase transitions, dubbed deconfined thermal transitions.
Constructing a $Z_2$ lattice gauge model with strong interactions between $Z_2$ gauge fluxes, we prove the existence of a thermal phase transition between deconfined and confined phases in two spatial dimensions in sharp contrast to its absence in the Wegner model. 
Incorporating deconfined fermions, it is shown that gapless excitations from Fermi surfaces endow line-tension to $Z_2$  gauge fluxes at zero temperature, and we argue that a deconfined thermal transition with deconfined fermions may be interpreted as a hidden order transition with thermal gap-opening in Fermi surfaces.     
Moreover, it is shown that symmetry breaking transitions in deconfined phases may be unconventional.  Global $Z_2$ and $U(1)$ symmetry breaking transitions in deconfined phases may be in the same universality class, which is impossible under the conventional Landau-Ginzburg-Wilson paradigm. 
Characteristic signatures of the transitions in experiments and candidate strongly correlated systems such as Kitaev materials are also discussed.  
 \end{abstract}

\maketitle

{\it Introduction :} 
Emergence of exotic excitations out of conventional electrons and spins is one of the striking characteristics of deconfined phases in strongly correlated systems \cite{Wen, Sachdev_topo}.  
Surprising quantum many-body effects such as the half quantization with a topological order may appear, and prime examples include Majorana fermions out of localized spins in quantum magnets \cite{Kitaev, Kitaev3d, Balents, Hermanns, Matsuda}.
It is quintessential to investigate deconfined phases and their propreties in strongly correlated quantum materials. 

Previous theoretical researches have mainly focused on quantum natures of deconfined phases including unconventional quantum phase transitions \cite{deconfined, Z2, Senthil1,Senthil2, MoonSachdev, You2012, MoonXu, Xu,Meng, Gazit, Grover, Gazit2, Hohenadler,You, duality, You2}. 
Properties of emergent particles are investigated,  and topological order beyond the Landau-Ginzbug-Wilson (LGW) paradigm is introduced. Also, unexpected duality relations between low energy theories around quantum phase transitions are unveiled.  

In this work, we consider thermal transitions associated with deconfined phases in two ($2d$) and three spatial dimensions ($3d$)  and demonstrate that deconfined phases even host exotic thermal phase transitions, named deconfined thermal transitions. One seminal work was done by Wegner who showed the presence of a thermal phase transition between confined and deconfined phases with the pure $Z_2$ lattice gauge model in 3d \cite{Wegner}. The transition has the aspects of a {\it hidden-order} transition because symmetry order parameters are absent in contrast to the conventional LGW paradigm. 
Below, we prove the presence of a thermal phase transition between confined and deconfined phases in $2d$ by constructing and analyzing a lattice model. Incorporating deconfined fermions, we also investigate thermal phase transitions between conventional and deconfined metals. 

We are also partly motivated by recent experiments  including exotic onset behaviors of order parameters in cuprates, iridates, and heavy fermions \cite{Matsuda1, Hsieh, Matsuda2, diagonal}. 
Inspired by these studies, novel universality class of thermal transitions breaking $Z_2$ and $U(1)$ symmetries associated with deconfined phases are obtained. Evaluating all critical exponents, we provide characteristic signatures of the transitions and candidate strongly correlated systems  in experiments.

{\it Pure $Z_2$ Lattice Gauge Theory  :} 
We first recall the Wegner model \cite{Wegner}, 
\begin{eqnarray}
H_W = - g \sum_{ \square} \prod_{(ab) \in \square} \sigma_{ab} \equiv -g \sum_{i^*} F_{i^*}
\end{eqnarray}
with $Z_2$ variables $\{  \sigma_{ab} = \pm1 \}$ on the link $(ab)$ between $a$ and $b$ sites. The $\square$ is for a plaquette whose position is specified by the dual index $i^*$. The gauge flux operator $F_{i^*}$ is introduced with a positive $g$. 
The Hamiltonian $H_W$ has two phases in 3d which was shown by mapping the model to the classical Ising model with the Kramer-Wannier duality transformation. At low temperatures ($g \gg T$), $Z_2$ gauge fluxes are frozen with the perimeter law of the Wilson-loop operator, but at high temperatures ($g \ll T$), $Z_2$ gauge fluxes are proliferated with the area law of the Wilson-loop operator, which are called deconfined and confined phases respectively. 
In the deconfined phase, it is useful to consider topological defects, $Z_2$ gauge flux loops, whose free energy is simply estimated to be $f_{3d} (l) \sim g \, l - T \log \Omega (l)$ with the length of the loop $l$. The entropy of the configuration $\Omega(l)$ shows power law dependence, and the transition temperature is an order of $T_* \sim g$. 

The absence of a deconfined phase in $2d$ may be understood by a similar estimation. A topological defect is not a loop but a point in $2d$, so the energy of a defect costs $2g$. The free energy of the two fluxes may be estimated as $f_{2d} (l) \sim 4g- T \log(\Omega(l))$ with the distance between the two fluxes, $l$. Therefore, at any non-zero temperatures, the entropic contribution wins and the gauge fluxes prefer to be proliferated, which disallows deconfined phases. Note that the same argument applies to the absence of superconductivity in 2d \cite{KT}. 

Explicit analysis on the model with a gauge choice, $\sigma_{i, i+\hat{x}}=1$, confirms the estimation.  
One can define a ``spin'' variable $S_i \equiv \sigma_{i, i+\hat{y}}$, and the flux operator becomes  $F_{i^*} = S_i  S_{i+\hat{x}}$. The unit vectors ($\hat{x}, \hat{y}$) on a square lattice are introduced. The Wegner model becomes equivalent to the decoupled set of  the one dimensional (1d) Ising models,
\begin{eqnarray} 
H_{W} = -g \sum_i S_i S_{i+\hat{x}}. \nonumber
\end{eqnarray}
and thus a thermal transition is absent at non-zero temperatures \cite{Kogut}. A domain-wall in each spin chain costs  finite energy, so its entropic contribution always wins proliferating domain-walls. 

We consider a lattice model of the spins, 
\begin{eqnarray}
H_{X} = -g \sum_i S_i S_{i+\hat{x}}  -J_r \sum_{i} \sum_{r =1} \frac{S_i S_{i+r\hat{x}}}{r^{\omega}}.
\end{eqnarray}
The term with $J_r$ describes the decoupled set of the Dyson-Ising chains \cite{Dyson}. The existence of a thermal phase transitions in the Dyson-Ising model was proven for $1< \omega \le 2$ \cite{Dyson, Spencer}, and the universality class is reported to be in the mean field class for $1< \omega < 3/2$ \cite{Blote}. 
The long-range interaction makes the domain-wall energy size-dependent, so its free energy may be estimated as $f(l) \sim J_r l^{2-\omega} - T \log \Omega (l)$.
At low temperatures, the flux is frozen, foramlly $\langle S_i \rangle \neq 0$, and the perimeter law of a thermal deconfined phase manifests as shown in Fig. 1.
One can readily replace the spins with gauge fluxes, and the model becomes
\begin{eqnarray}
 H_{X} = - g \sum_{i} F_{i^*}-J_r \sum_{i} \sum_{r=1 } \frac{\prod_{a=0}^{r-1}F_{i^*+a\hat{x}}}{r^{\omega}},
\end{eqnarray}
which may be further generalized to  
\begin{eqnarray}
&&H_Z = - \sum_{i} \sum_{r=1 } J(r) \prod_{a=0}^{r-1}F_{i^*+a\hat{x}}.
\end{eqnarray}
Defining the two constants $M_0 \equiv \sum_{r=1}^{\infty} J(r)$ and  $K_3^{'} = \sum_{r=1}^{\infty} \big( {\rm log \, log} (r+4) \big) [r^3 J(r)]^{-1}$ with monotonically decreasing $J(r) \ge 0$, we can map $H_Z$ to the decoupled set of the Dyson-Ising chains, which guarantees the existence of a deconfined thermal phase and its transition to a confined phase for finite $M_0$ and $K_3^{'}$\cite{Dyson}.

We stress that the strong interaction between the gauge fluxes is the impetus of a thermal deconfined phase in $2d$ in sharp contrast to the Wegner model. 
Our model breaks a rotational symmetry by picking up the $\hat{x}$ direction, yet we believe that the symmetrized model, $H_{\bar{Z}} = (H_Z ( \hat{x})+H_Z ( \hat{y}))/2 $, has the presence of the deconfined phase. More realistic flux models to realize the thermal deconfined phases  with numerical analysis will be presented in future works. 
The Wegner model is a non-interacting theory of gauge fluxes upto the gauge constraint, and our model shows that a strongly interacting gauge flux theory realizes a deconfined phase even in 2d.   

\begin{figure}
\includegraphics[width=3.2in]{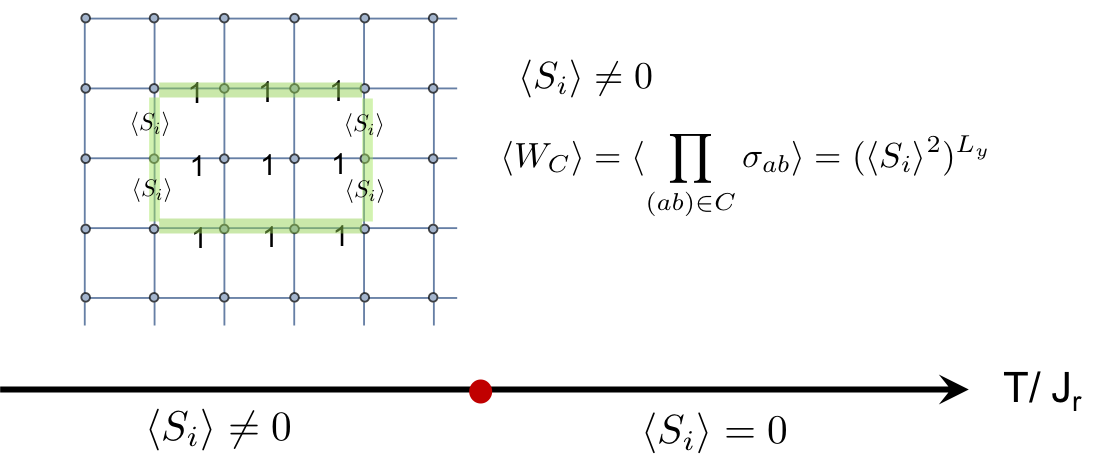}
\caption{ Perimeter law of the low temperature phase of $H_{Z}$. At low temperatures, a non-zero value of $\langle S_i \rangle$ indicates the perimeter law of the Wilson loop operator ($W_C$). 
}
 \end{figure}

{\it Models with deconfined fermions :} 
We extend the model by incorporating fermions ($f_{i}$) and Ising spins ($s_i=\pm1$) on the sites of a hyper-cubic lattice,
\begin{eqnarray}
H_{\sigma f} = H_{Z} -  \sum_{\langle i,j \rangle} J_{ij} \sigma_{ij} s_{i} s_j  -  t \sum_{\langle ij \rangle} \sigma_{ij} f^{\dagger}_{i} f_{ j } +V_f.\nonumber
\end{eqnarray}
The $Z_2$ gauge structure is manifested by a gauge transformation,
$\sigma_{ij}  \rightarrow \sigma_{ij} \eta_i \eta_j$, $s_i \rightarrow \eta_i s_i$, and $f_{i \alpha} \rightarrow f_{i \alpha} \eta_i$  with $\eta_i=\pm1$. 
A gauge invariant potential of fermions, $V_f$, is introduced. The model with $t=0$ and $J(r) = g \delta_{r=1}$ was considered by Fradkin and Shenker \cite{Fradkin}, showing the equivalence between the Higgs phase of $s_i$ and the confined phase in $3d$. 

Quantum mechanical analysis is necessary for fermions and one can treat $\{\sigma_{ij}, s_i \}$ as static background fields.  
In the deconfined phases in $2d$ and $3d$, one can safely ignore the Ising spins, and the ground state energy with the zero gauge flux $\{ \sigma^0_{ij} \}$ is obtained by diagonalizing the fermion Hamiltonian, 
\begin{eqnarray}
H_f(\{  \sigma^{0}_{ij} \} ) = -  t \sum_{\langle ij \rangle} \sigma_{ij}^0 f^{\dagger}_{i} f_{ j } +V_f, \nonumber
\end{eqnarray}
and filling up fermions to a chemical potential. The Hamiltonian with the two gauge fluxes ($\{ \sigma^{2F}_{ij} \}$)  is
\begin{eqnarray}
 H_f(\{  \sigma^{2F}_{ij} \} ) =H_f(\{  \sigma^{0}_{ij} \} )  -t \sum_{\langle i j \rangle}(\sigma^{2F}_{ij}- \sigma^{0}_{ij}) f^{\dagger}_{i \alpha} f_{j \alpha}. \nonumber
\end{eqnarray}
It is convenient to choose the gauge $\{ \sigma^0_{ij}=1 \}$ on every link and $\{ \sigma^{2F}_{ij} \}$ differs from $\{ \sigma^0_{ij} \}$ only in the interconnecting line between the two fluxes. 
Without loss of generality, we may assume that the two fluxes are separated along the $x$ direction whose distance is $l_{2F}$. It is useful to notice that 
the second term on the right hand side is a perturbation to the first term and one can perform the perturbative calculation with a small parameter $l_{2F}/N_{size}$ with the system size $N_{size}$ setting a lattice constant as a unit.
For simplicity, let us consider the non-interacting limit $V_f \rightarrow 0$ and diagonalize the Hamiltonian exactly, and  
the energy difference at the leading order is    
 \begin{eqnarray}
 E_{f0}(\{  \sigma^{2F}_{ij} \}) -  {E}_{f0}(\{ \sigma^0_{ij} \}) =    \Big[\frac{t}{N_{site}} \sum_q n_F(q)( 2\cos( q_y))\Big] l_{2F}  \nonumber
\end{eqnarray}
with the Fermi-Dirac function $n_F(q)$ in 2d. Its 3d generalization is straightforward.  We may assign the right hand side to the tension energy between the two fluxes. It is easy to show that the second term is positive because the summation range is determined by the sign of $t$. 
Since the line-tension only depends on the particle number and quasi-particle dispersion relations, we believe the calculation is perturbatively safe.

Our calculations indicate that the Fermi surfaces may be a source to stabilize thermal deconfined phases by providing interaction channels between fluxes. 
Strictly speaking, our line-tension calculations are done at $T=0$, and it is desired to check temperature dependence of the line-tension.  
Note that the interaction between the fluxes has a formal similarity to the Ruderman-Kittel-Kasuya-Yoshida interaction \cite{RKKY} and 
a power counting of the interaction shows that the Fermi surfaces induce a similar type of the long-range interaction, which is also desired to be checked by numerical and analytical calculations in future works.
 
At high temperatures, entropic contribution of topological defects dominate and a confined phase appears which may be understood as a Higgs phase of the Ising spin. Gauge-neutral fermions, $c_{i\alpha} \equiv s_i f_{i \alpha}$, become good degrees of freedom, where Fermi liquids of $c$ fermions are expected. It is a conventional metal distinguished from a deconfined metal of $f$ fermions at low temperatures.
 
Let us consider the critical theory under the fermionic fluctuations. The critical theory in 2d will be discussed in a future work, and in 3d, without the $f$ fermions, the transition of the Wegner model is the dual-Ising class with the dual Ising variable $\zeta(x)$ which is a {\it trivial} representation of all symmetries. The variable is coupled to the number density of fermions at the lowest order \cite{Senthil2}, and the critical theory is
 \begin{eqnarray}
\mathcal{S}_{c} &=&  \int d^3 x \Big[  \frac{1}{2}(\nabla \zeta(x))^2 + \frac{r}{2} \zeta(x)^2+\frac{\lambda}{4!} \zeta(x)^4 \Big] \nonumber\\ 
 &+& \int d^3 x d \tau \Big[ f^{\dagger}(x,\tau)\partial_{\tau} f_(x,\tau) +\mathcal{H}_f(x,\tau) \Big]  \nonumber  \\
 &+& \int d^3 x d \tau  (- g_2 \zeta(x)^2  - g_4 \zeta(x)^4 )n_f(x,\tau) +\cdots \nonumber
\end{eqnarray}
 with $n_f (x, \tau) \equiv f^{\dagger}(x,\tau)f(x,\tau)$ and the fermion Hamiltonian density, $\mathcal{H}_f$. 
 The tuning parameter has the temperature dependence, $r \propto T_*-T$, with the transition temperature $T_*$. Note that the sign of temperature dependence is opposite to conventional phase transitions. 
 
 In contrast to the dual Ising variable, fermions explicitly depend on  imaginary time reflecting their quantum natures as usual. 
Defining the density fluctuation,  $\delta n_f(x,\tau) \equiv n_f(x,\tau) - \bar{n}_f$ with  $\bar{n}_f \equiv \frac{N_f}{N_{site}}$ and the total fermion number $N_f$, we integrate out the fermions.
Evaluating the Yukawa term over the fermion path-integral,
 \begin{eqnarray}
 &&\langle e^{+\int d^3 x  \big(g_2\zeta(x)^2 + g_4 \zeta(x)^4 \big)\delta n_f(x,\Omega_n=0)} \rangle_f,\nonumber 
 \end{eqnarray}
  the coupling constants are modified as
\begin{eqnarray}
r \rightarrow r -g_2 \frac{\bar{n}_f}{T}, \quad \lambda \rightarrow \lambda - g_4 \frac{\bar{n}_f}{T}, \nonumber
\end{eqnarray} 
(see SI).  We use $\delta n_f(x,\Omega_n=0) = \int d\tau \delta n_f(x,\tau)$. It is important to notice the absence of additional singular contributions from thermal fermionic fluctuations.  For small enough coupling constants ($g_2, g_4, \cdots$), the phase transition is in the dual Ising class. But for a large enough $g_4$, the self-interacting term of the dual Ising field may become negative with a large enough $g_4$ signaling a first order transition. 

A deconfined metal is one concrete realization of the orthogonal metal proposed in the seminal work by Nandkishore, Metlitski, and Senthil \cite{Senthil2}. 
We remark few points. First, the orthogonal metal to conventional metal transition at zero temperature is shown to be {\it not} of the dual Ising class because fermionic fluctuations are too strong. In thermal transitions, fermionic fluctuations may not strong enough and a second order transition of the dual Ising class may be realized. Second, the thermal transitions indicate that the correlation functions of the $c_{i\alpha}$ fermions are suppressed below  $T_*$.  The Fermi surfaces of $c$ fermions may lose spectral weights below $T_{*}$ similar to the onset of the gap onset $\Delta \sim \xi^{-z} \sim (T_*-T)^{z\nu}$ if the transition is continuous. This may be observed in angle-resolved-photo-emmision-sepctroscopy (ARPES) experiments. The dynamic critical exponent is expected to be $z=1$ for the dual Ising field if there are no other dynamic channels. Thus, specific heat and ARPES have definite signatures of the transition at $T_*$ while other static experiments with charge or spin degrees of freedom are featureless. Therefore, a deconfined thermal transition with deconfined fermions may be interpreted as a hidden order transition with thermal gap-opening in Fermi surfaces.

\begin{table}[tb!]
\begin{tabular} {|c|c|c|c|c|c|c|}
\hline \hline
Univ. class in $3d$ & $\alpha$  & ~~$\beta$~~  & ~~$\gamma$~~ & ~~$\nu$~~ & ~~$\eta$~~ & ~~$\delta$~~  \\ \hline \hline
$Z_2$ (Ising) \cite{Vicari_RG} & $0.11$ & $0.33$ & $1.24$ & $0.63$ & $0.036$ & $4.79$  \\ \hline  
$U(1)$ (XY) \cite{Vicari_RG} & $-0.015$ & $0.35$ & $1.32$ & $0.67$ & $0.038$ & $4.78$ \\ \hline \hline 
Mean-field   & $0$ & $0.5$ & $1$ & $0.5$ & $0$ & $3$  \\ \hline  
 \hline
DC-$Z_N$/DC-U(1)   & $-0.015$    & $0.83$ & $0.35$ & $0.67$ & $1.47$& $1.43$  \\ \hline \hline
\end{tabular}
\caption{Universality classes in $3d$.  The first raw is for the critical exponents ($C \sim |t|^{-\alpha}$, $\langle \Phi \rangle \sim |t|^{\beta}$, $\chi  \sim |t|^{-\gamma}$, $\xi^{-1} \sim |t|^\nu$, $[\Phi] = \frac{1+\eta}{2}$, and $\langle \Phi \rangle \sim h^{1/\delta}$ ) with an order parameter $\Phi$ and conjugate field $h$. A reduced temperature $t \equiv (T-T_c)/T_c$ is used. The DC-$Z_N$ and DC-$U(1)$ are for the universality classes of the deconfined thermal phase transitions with $Z_N$ and $U(1)$ symmetries, respectively. Note that DC-$Z_2$ and DC-$U(1)$ are in the same universality class in sharp contrast to the $Z_2$ and $U(1)$ transitions under the LGW paradigm (see also SI for larger gauge groups).
}
\end{table}

{\it Symmetry Breaking Transitions : }
Symmetry breaking transitions associated with deconfined phases may be different from the ones in confined phases. 
Since fermionic fluctuations and bosonic ones with non-zero Matsubara frequencies are irrelevant to a critical theory\cite{Sachdev_QPT}, one can focus on the static component of the bosonic fluctuations. 
 
To be specific, let us consider the case with a global symmetry $Z_N$. One conventional way to represent $Z_N$ is to introduce an angle variable ($\theta_i$) of the $2\pi$ periodicity with the potential term, 
$V_b(\{ b_i\}) = -u \sum_{i} \cos(N \theta_i)$.
For $u>0$, the $N$ minimal configurations are $\theta_i = \frac{2\pi}{N} n_i$ with $n=0,\cdots, N-1$, and the order parameter is $\Phi_i\propto (\cos(\theta_i), \sin(\theta_i)) $. 
The Landau theory with the $Z_N$ symmetry is  
$F_{L} = - \tilde{J} \sum_{i,j} \cos(\theta_i-\theta_j) -u \sum_{i} \cos(N \theta_i)$,
 which is  the conventional $Z_N$ clock model. 

To go beyond the LGW paradigm, let us consider the Hamiltonian,
\begin{eqnarray}
H_s = -J \sum_{\langle ij \rangle} \sigma_{ ij  }\cos(\frac{\theta_i - \theta_j}{2}) -u \sum_{i} \cos(N \theta_i) - g \sum_{i} F_{i^*}. \nonumber
\end{eqnarray}
The local gauge transformation includes $\theta_i \rightarrow \theta_i +2\pi$ and $\sigma_{ij} \rightarrow -\sigma_{ij}$ for all links at  $i$ site.  
The confined phase ($g/T \ll1$) may be studied by using the high temperature expansion, and  one can obtain $F_L$ with higher order terms by selecting gauge-invariant terms.

In the deconfined phase ($g/T \gg1$) (or $J_r /T \gg1$ in 2d),  the gauge flux is frozen with ($\sigma^0_{ij}=1$), and the effective Hamiltonian becomes
\begin{eqnarray}
H_s = - J \sum\cos(\frac{\theta_i-\theta_j}{2}) -u \sum_{i} \cos(N \theta_i), \nonumber
\end{eqnarray}
upto an unimportant constant. 
We stress that the periodicity of $\theta_i$ becomes  $4\pi$ in the deconfined phase. 
This is because the $2\pi$ periodicity of $\theta_i$ should be accompanied by the gauge transformation but  the gauge flux is frozen prohibiting the gauge transformation with $\theta_i \rightarrow \theta_i +2\pi$. 
In other words, the $2\pi$ vortex configuration of $\theta_i$ is confined, and only the $4\pi$ vortex configuration is allowed.

Introducing the half angle variable $\vartheta_i = \frac{\theta_i}{2}$, the Hamiltonian and order parameter are rewritten as 
\begin{eqnarray}
H_s = -J \sum\cos( \vartheta_i-\vartheta_j) -u \sum_{i} \cos(2N \vartheta_i) 
\end{eqnarray}
and $\Phi_i \propto (\cos(2\vartheta_i), \sin(2\vartheta_i))$. Note that the half angle operator $\langle e^{i \vartheta_i} \rangle $ carries the $Z_2$ gauge charge, so it vanishes by definition. 
Thus, the $Z_N$ symmetry breaking transition in the deconfined phase is described by the $Z_{2N}$ clock model whose order parameter ($\langle \Phi_i \rangle$) is a secondary operator of $\vartheta_i$.

In 2d, this is precisely mapped to the recent proposal of the inverted clock model universality class with central charge one even for $Z_2$ symmetry breaking transitions \cite{Moon1}.  
Its critical theory in 3d may be conveniently expressed by introducing a gauge charged field, $\vec{\phi}\equiv (\phi_x, \phi_y)= \rho_0 (\cos(\vartheta),\sin(\vartheta) )$ restoring the amplitude mode $\rho_0$,
\begin{eqnarray}
S_{DC} = \int d^3 x (\nabla \vec{ \phi})^2 + r (\vec{\phi})^2 + \frac{\lambda}{4} ((\vec{\phi})^2)^2 - \tilde{u} (\phi_x^{2} - \phi_y^{2} )^N. \nonumber
\end{eqnarray}
The gauge-neutral order parameter is $\vec{\Phi} = (\phi_x^2-\phi_y^2, 2 \phi_x \phi_y)$ since $\vec{\phi} \rightarrow - \vec{\phi}$ under the gauge transformation. 
The anisotropy term with $\tilde{u}$ is well understood in literature, which is irrelevant for $N \ge 2$ to the $U(1)$ fixed point \cite{Vicari}. 
In other words, the universality classes of $Z_N$ and U(1) symmetry breaking transitions are the same. 
The tuning parameter critical exponent ($\nu$) is obtained from  $(\vec{\phi})^2$, which is known as $\nu=0.67155$ \cite{Vicari}.
The scaling dimension of the order prameter is known to be $[\vec{\Phi}] =1.24$ \cite{Vicari}, and thus the order parameter onset exponent $\beta= \beta_2 \equiv 0.83$. 
With the two independent exponents and the scaling relations, we find all the critical exponents summarized in Table I. 
Note that the negative value of $\alpha$ indicates that the Harris criteria is valid under quenched disorder. 

Few remarks are as follows. Generalization to other symmetry groups is straightforward similar to a nematic phase adjacent to a deconfined phase \cite{Toner, Beekman}. In $2d$ at $T=0$, the enlarged periodicity has also been discussed in the context of quantum phase transitions in frustrated quantum magnets \cite{Huh}. The quantum-classical mapping connects the DC-$U(1)$ class  with  quantum XY$^*$ class \cite{Huh,Senthil5}. 
Thus, our lattice model analysis shows that $Z_2$ symmetry potential terms may be irrelevant to the quantum XY$^*$ class, which describes a subset of spin-exchange interactions in frustrated quantum magnets. Moreover, our results can be easily extended to higher gauge groups. For example, order parameters are higher order operators whose onset critical exponents are $\beta_3 = 1.42$ and $\beta_4 = 2.09$ for $Z_3$ and $Z_4$ gauge groups respectively \cite{Vicari}. Since the critical exponent $\alpha$ is the same, all the critical exponents are determined with $\alpha$ and $\beta_{M\ge3}$ (see SI). 
 
\begin{figure}
\includegraphics[width=3.2in]{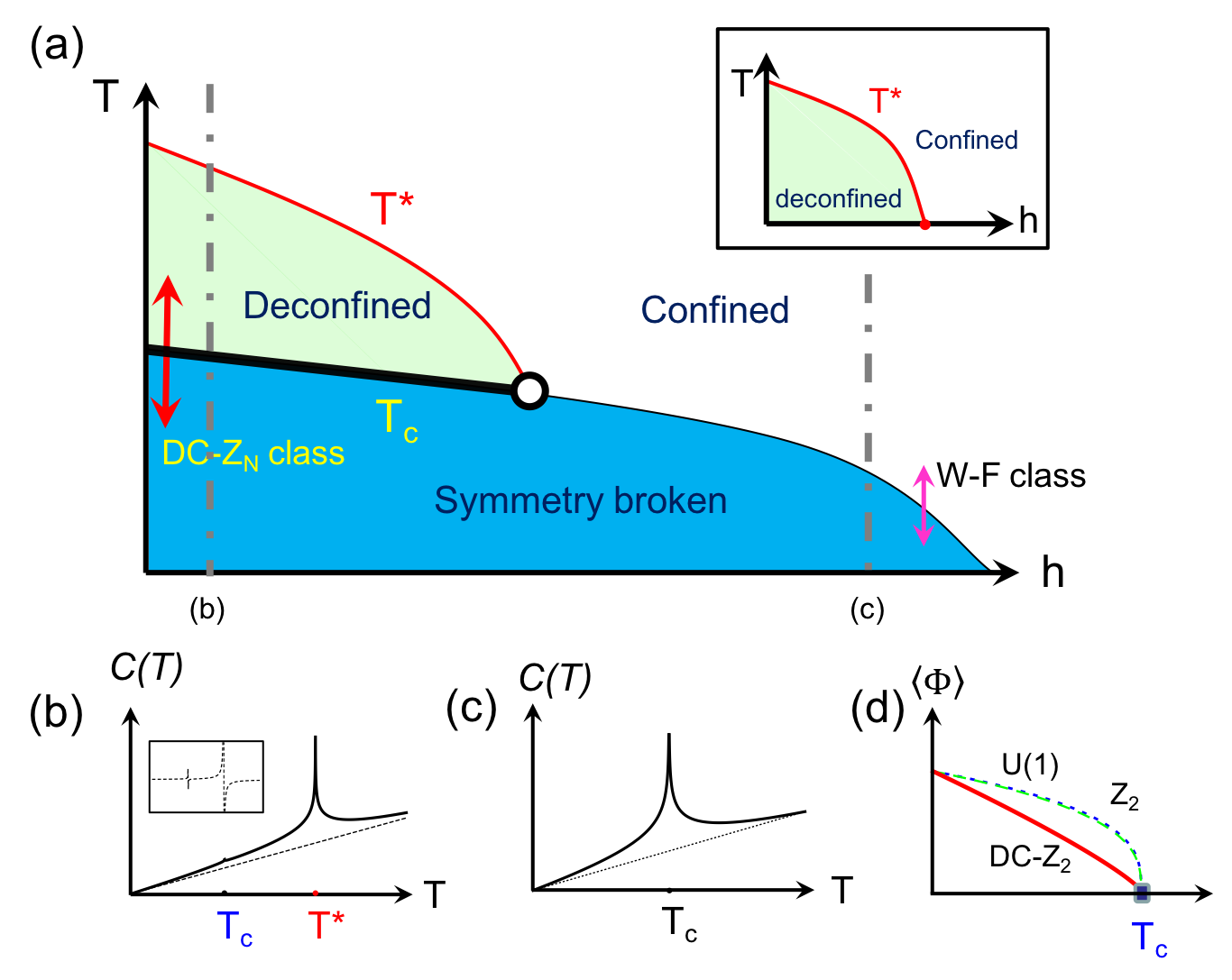}
\caption{ (a) Schematic phase diagram associated with a deconfined thermal phase in 3d. The two dimensionless parameters $T/g$ and $h/g$ are introduces, which characterize temperature and quantum fluctuations, respectively.  Phase transition between confined and deconfined metals at $T_*$ without breaking any symmetries. The thin black line of a symmetry breaking transition is associated with the conventional Wilson-Fisher universality class. The thick black line is associated with our exotic universality class, DC-$Z_N$/$U(1)$. The inset is for a phase diagram without breaking symmetries. (b) Specific heat $C_v(T) = a T + C_{sing}(T)$ associated with the deconfined thermal phase. The two transitions at $T=T_c$ and $T=T^*$ are associated with a symmetry breaking and flux-line proliferation. The constant $a$ is for background metallic contributions. The inset is for $d C_v / dT$, and it is obvious that the singularity of the transition at $T_c$ is weaker than the one of $T^*$.
(c) Specific heat $C_v(T) = a T + C_{sing}(T)$ around a conventional $Z_2$ symmetry breaking transition without the deconfined thermal phase.
(d) The order parameter ($\langle \Phi \rangle$) onsets below $T_c$. We find $\beta = 0.83$ for the DC-$Z_N$/$U(1)$ classes. The  conventional $Z_2$ and $U(1)$ classes have  $\beta_{Z_2} =0.33$ and $\beta_{U(1)}=0.35$ respectively. 
}
 \end{figure}

 {\it Discussion and Conclusion :} 
A schematic phase diagram associated with a thermal deconfined  is illustrated in Fig. 2. 
The universality class of symmetry breaking transitions associated with a deconfined thermal phase is not the conventional Wilson-Fisher fixed class but the deconfined $Z_N$/$U(1)$ class, which have striking characteristics. For example, a $Z_2$ symmetry breaking transition under the Landau paradigm shows significant specific heat anomaly such as jumps or divergences. On the other hand, DC-$Z_2$ has the negative value of $\alpha$, so specific heat anomaly is less significant in experiments.  Moreover, the order parameter onset is much slower below $T_c$ as manifested in $\beta_M$.  One non-trivial consequence of our new universality class is that specific heat shows a bigger anomaly at higher temperature, $T^*$, which may be considered as a ``hidden order'' transition. Our results provide one explanation of hidden order transitions in the absence of broken symmetries. For example, one recent experiment in a naturally hetero-structured system, Sr$_2$VO$_3$FeAs \cite{JSKim}, reported a thermal phase transition without breaking any symmetries.
 
Note that the recently proposed Kitaev materials \cite{ Hermanns} may be a promising platform since the low temperature state is already the deconfined quantum spin liquids with Majorana excitations \cite{Motome}. We believe symmetry breaking phenomena in Kitaev materials may show unconventional behaviors.

In conclusion, deconfined thermal phase transitions are demonstrated. 
We prove the existence of a thermal phase transition in 2d with $Z_2$ gauge fields and thermal phase transitions between conventional and deconfined metals are illustrated. 
Unconventional symmetry breaking transitions in deconfined phases are also presented. Namely, the $Z_2$ and $U(1)$ symmetry breaking transitions in 3d are in the same universality class which is impossible under the LGW paradigm. 
Our results may be generalized and applied to other topological phases such as exotic phases with fracton excitations. 
Future studies on numerical tests  incorporating quantum fluctuations of the $Z_2$ gauge fields would be useful, and detailed studies on relations with microscopic models and experiments such as doped Kitaev materials and heavy fermions are highly desired.

 {\it Acknowledgement :} 
 We thank Leon Balents, Eduardo Fradkin,  Yong Baek Kim, Steven Kivelson, Sungjay Lee, Max Metlitski, Subir Sachdev, and Cenke Xu for invaluable discussions and comments.  
This work was supported by the POSCO Science Fellowship of POSCO TJ Park Foundation and NRF of Korea under Grant No. 2017R1C1B2009176.

\bibliographystyle{apsrev4-1}
\bibliography{DCthermal}

\newpage
 \appendix

\section{Comments on $H_Z$}

We make comments on  the generalized model, 
\begin{eqnarray}
&&H_Z = - \sum_{i} \sum_{r=1 } J(r) \prod_{a=0}^{r-1}F_{i^*+a\hat{x}} 
\end{eqnarray}
with $M_0 \equiv \sum_{r=1}^{\infty} J(r)$ and  $K_3^{'} = \sum_{r=1}^{\infty} \big( {\rm log \, log} (r+4) \big) [r^3 J(r)]^{-1}$ for $J(r) \ge 0$. 
The model has a thermal phase transition for $1< \omega \le 2$. One can apply the Dyson's theorem for $1<\omega <2$ but for $\omega=2$, the existence is out of the Dyson's theorem. Yet, the $\omega=2$ case is also proven \cite{Spencer}.  
The constant infinite range  interaction ($\omega=0$) does not belong to the phase transition criteria because the domain-wall energy diverges in the thermodynamic limit. For $\omega>2$, the domain-wall energy becomes finite, so the model becomes adiabatically connected to the Wegner model.
 
We note that the specific form of the interaction in $H_Z$ is used to prove the existence of a deconfined thermal phase and its transition. It is highly desired to find a simpler model with a short range interaction, for example  
\begin{eqnarray}
H_{m Z} = - g \sum_i F_{i^*} - \sum_{\langle i, j \rangle} J_{ij} F_{i^*} F_{j^*},  
\end{eqnarray}
with an ``exchange'' interaction $J_{ij}$. It is highly desired to find a lower bound of the effective range of $J_{ij}$ which realize a thermal deconfined phase.

\section{Thermal phase transition of $H_Z$ with Ising matter field }
Let us consider the model, 
\begin{eqnarray}
H &=& - \sum_{i} \sum_{r=1 } J(r) \prod_{a=0}^{r-1}F_{i^*+a\hat{x}}  - J_{FS} \sum_{\langle ij \rangle} \sigma_{ij} \phi_i \phi_j \nonumber \\
&=&  -J_r \sum_{i} \sum_{r =1} \frac{S_i S_{i+r\hat{x}}}{r^{\omega}} - J_{FS} \sum_{i} (\phi_i \phi_{i+\hat{x}}  +S_i \phi_i \phi_{i+\hat{y}} ). \nonumber
\end{eqnarray}
In the second line, we choose the gauge $\sigma_{i i+\hat{x}}=1$ and $\sigma_{i i+\hat{y}} = S_i$, and the Hamiltonian is described by the two types of spins $S_i = \pm1$ and $\phi_i = \pm1$. 

Starting with the presence of the phase transition with $J_{FS}=0$, we investigate effects of $J_{FS}$ perturbatively. 
Let us consider the regime $J_{r},T \gg J_{FS}$ where the high temperature expansion with $J_{FS}/T$ is possible. 
The partition function is 
\begin{eqnarray}
Z &=& {\rm Tr} ( e^{-H/T}) \nonumber \\
&=& \sum_{\{S_i \}, { \{\phi_i\}}} \prod_{i} e^{\sum_{r =1} \frac{J_r}{T} \frac{S_{i} S_{i+r\hat{x}}}{r^{\omega}}}e^{\frac{J_{FS}}{T} \phi_i \phi_{i+\hat{x}}}  e^{\frac{J_{FS}}{T} S_i \phi_i \phi_{i+\hat{y}}   }\nonumber \\
&\equiv&\sum_{\{S_i \} } \prod_{i} e^{\sum_{r =1} \frac{J_r}{T} \frac{S_{i} S_{i+r\hat{x}}}{r^{\omega}}} G(\{ S_i\}).
\end{eqnarray}
The function $G(\{ S_i\})$ may be obtained by 
\begin{eqnarray}
G(\{ S_i\}) &=& \sum_{\{\phi_i \}} \prod_{i}  e^{\frac{J_{FS}}{T} \phi_i \phi_{i+\hat{x}}} \Big(1 + \tanh(\frac{J_{FS}}{T}) S_i \phi_i \phi_{i+\hat{y}} \Big). \nonumber \\
&=&1+ \sum_{i, r=1} a_{r} S_{i} S_{i+r \hat{x}}  + \sum_{i,j,k,l} b_{ijkl} S_i S_j S_k S_l +\cdots \nonumber
 \end{eqnarray}
 upto unimportant constants.
By using the Taylor expansion, one can show the coefficients have the exponential behaviors, 
\begin{eqnarray}
a(r) \propto (\frac{J_{FS}}{T})^{r} = e^{-r /\xi_{FS}}, \quad \xi_{FS}^{-1} \equiv \log(\frac{T}{J_{FS}}), \nonumber
\end{eqnarray}
and the higher order terms also show the exponential decay, and the correlation length is tiny for $J_{FS}/T \ll1$. 
Assuming a second order transition for $J_{FS}=0$, we may use the argument by Fradkin and Shenker  \cite{Fradkin}. Namely, the short-range interactions with $J_{FS} \neq0 $ are unable to destabilize the presence of the transition.  
We argue that the transition is perturbatively stable.

The self-consistency may be checked as follows. At the temperature regime $J_r \gg T \gg J_{FS}$, one may set $\langle S_i \rangle \neq 0$ formally. By using a gauge transformation, one can make the expectation value uniform positive. Then, the effective Hamiltonian of $\phi_i$ becomes the anisotropic Ising model whose critical temperature is determined by 
\begin{eqnarray}
\sinh(\frac{2 J_{FS}}{T_c}) \sinh(\frac{2 J_{FS} \langle S_i \rangle}{T_c})=1. \nonumber
\end{eqnarray}
This demonstrates the stability of the flux frozen thermal phase  for $J_{r} \gg T \gg J_{FS}$. At high temperatures $T \gg J_{r} \gg J_{FS}$, the flux becomes proliferated, so there is a phase transition between the flux frozen and flux proliferated phases.    

Note that the matter field carries the $Z_2$ electric charge but not the dual $Z_2$ magnetic charge. Thus, the phase transition with the $Z_2$ electric charged particles do not qualitatively modify the thermal phase transition \cite{Senthil1}.

\section{Line-tension calculation}
In this section, we provide details of the line-tension calculation between the two fluxes.
Without loss of generality, we assume that the two fluxes are separated along the $x$ direction as in Fig. \eqref{visons}.
The fermionic Hamiltonian with the two fluxes ($\{ \sigma^{2F}_{ij} \}$) may be rewritten as
\begin{eqnarray}
 H_f(\{  \sigma^{2F}_{ij} \} ) =H_f(\{  \sigma^{0}_{ij} \} )  - t \sum_{\langle i j \rangle} (\sigma_{ij}^{2F} - \sigma_{ij}^0) f^{\dagger}_{i} f_{j }. \label{flux}
\end{eqnarray}
We fix the gauge choice here after and the zero flux notation ($\{\sigma^0_{ij}=1\}$) is used. The second term is for the two flux states separated by the length $l_{2F}$. The notation $\langle i j \rangle  \in l_{2F}$ accounts for the interconnecting links (links with crosses). 
For simplicity, we set $V_f = 0$, which allows us to perform full analytic calculations. It is obvious that the second term on the right hand side is a perturbation to the first term. 
Introducing the Fourier transformation of the fermion variables, 
\begin{eqnarray}
f_j = \frac{1}{\sqrt{N_{site}}} \sum_q f_q e^{i q x_j}, \nonumber
\end{eqnarray} 
the second term becomes 
\begin{eqnarray}
2t \sum_{\langle i j \rangle  \in l_{2F}} f^{\dagger}_{i} f_{j } =  \frac{2t}{N_{site}} \sum_{k,q} f_k^{\dagger} f_q \sum_{j=1}^{l_{2F}} \cos(k_x j - q_x j + q_y). \nonumber
\end{eqnarray}
The ground state of the zero flux state $| G \rangle = \prod| k\rangle$ may be used to determine the estimation of the energy with the two fluxes, which becomes
\begin{eqnarray}
E_{0}(\{  \sigma^{2F}_{ij} \}) &=&  {E}_{0}(\{ \sigma^0_{ij} \}) + 4g + 2t \frac{l_{2F}}{N_{site}}  \sum_q n_F(q) \cos( q_y) \nonumber
\end{eqnarray}
with the Fermi-Dirac function $n_F(q)$.
Note that the positive sign of $t$ makes the summation range ($\sum_{|q| < k_F}$) from the Fermi-Dirac function, and thus the second term is always positive. The $k_F$ is determined by the particle number density of fermions, and a filled band vanishes the summation $\sum_{|q| < k_F} \cos(q_y)$. 
If we use the conventional Sommerfeld expansion, then the free-energy of the two fluxes with $l$ may be estimated as
\begin{eqnarray}
F_{2F} \sim \rho(k_F) t  \, l   -T \log \Omega (l),
\end{eqnarray}
where $\rho(k_F)$ is for the summation of $n_F(q) \cos(q_y)$ which is non-zero in the presence of the Fermi surfaces. 
The transition temperature would be estimated as $T_* \sim \rho(k_{2F}) t $ under our crude approximation, which needs to be carefully improved in future works.

\begin{figure}
\includegraphics[width=2.4in]{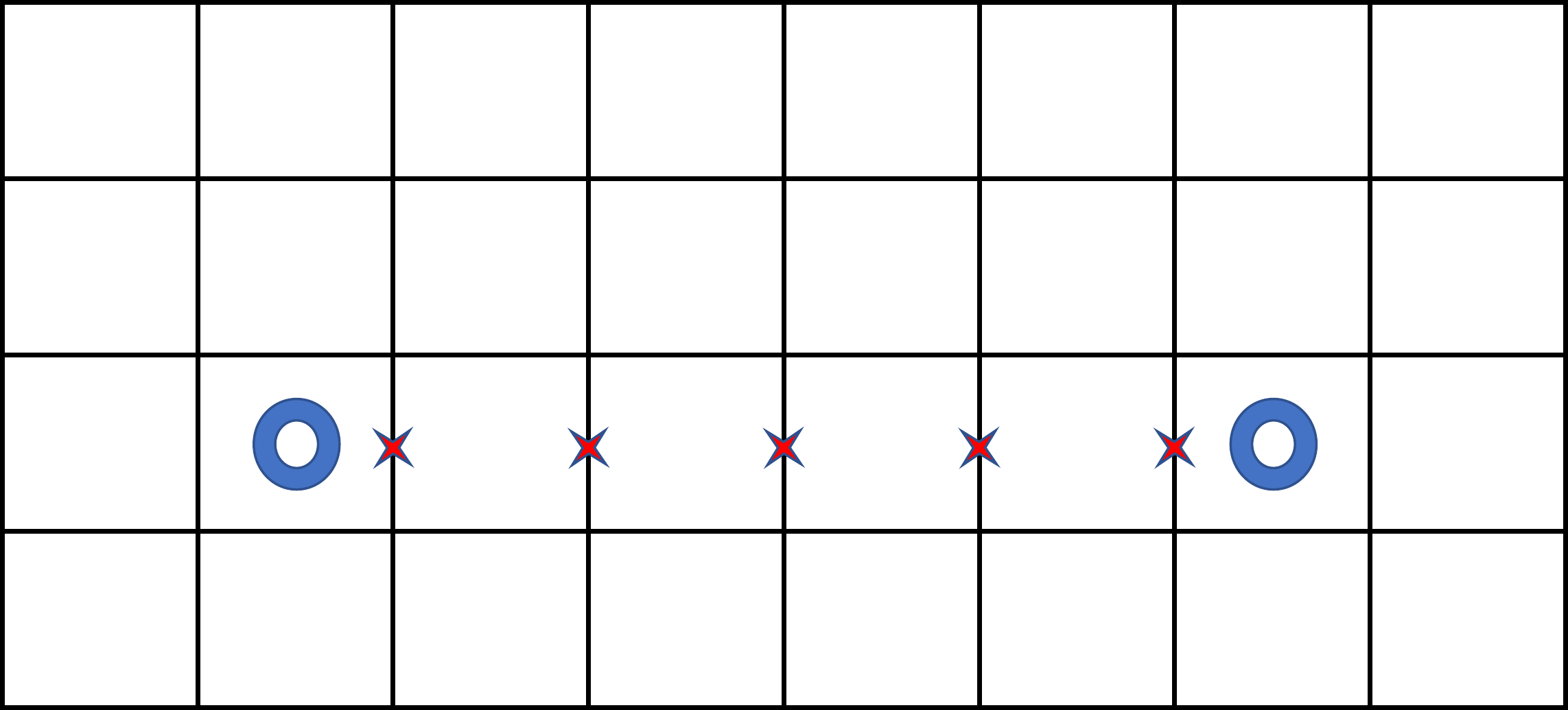}
\caption{The state with two gauge fluxes (circles). The two types of the links are shown : plain links ($\sigma_{ij}=1$) and the links with crosses ($\sigma_{ij}=-1$). 
The length between the two fluxes is $l_{2F}=5$ in the unit of the lattice spacing. 
} \label{visons}
 \end{figure}
 
 \section{Evaluation of the Yukawa coupling}
 
 The Yukawa type interaction term in the partition function is 
 \begin{eqnarray}
 &&\langle e^{\int d^3 x  ( g_2\zeta(x)^2 +g_4 \zeta(x)^4)\delta n_f(x,\Omega_n=0)} \rangle_f\nonumber \\
 &=& \sum_m \frac{1}{m} \langle \Big(\int d^3x (g_2 \zeta(x)^2 +g_4 \zeta(x)^4)\delta n_f(x,\Omega_n=0) \Big)^m\rangle_f \nonumber \\
 &=&1 + \int d^3 x ( g_2 \zeta^2(x) + g_4 \zeta^4(x) )\langle \delta n_f(x,\Omega_n=0) \rangle_f \nonumber \\
 &+&\frac{g_2^2}{2} \int d^3x d^3 y \zeta(x)^2 \zeta(y)^2 \langle \delta n_f(x,\Omega_n=0) \delta n_f(y,\Omega_n=0) \rangle_f \nonumber \\
 &+& \cdots
 \end{eqnarray}
The fermionic correlation functions may be easily obtained, 
\begin{eqnarray}
\langle \delta n_f(x,\Omega_n=0) \rangle_f = \int d\tau \langle \delta n_f(x,\tau) \rangle_f = 0, \nonumber
\end{eqnarray}
assuming the fermionic ground state does not break translational symmetry. 
In the term with $g_2^2$, we need the contact term, 
\begin{eqnarray}
 \langle \delta n_f(x,\Omega_n=0) \delta n_f(y,\Omega_n=0) \rangle_f  = c \delta^3 (x-y) + \cdots, \nonumber
\end{eqnarray} 
and one can determine $c$ by evaluating
\begin{eqnarray}
c &=& \int d^3x \langle \delta n_f(x,\Omega_n=0) \delta n_f(0,\Omega_n=0) \rangle_f \nonumber \\
&=& \int d^3x \int d\tau_1 \int d \tau_2 \langle \delta n_f(x,\tau_1) \delta n_f(0,\tau_2) \rangle_f = 0. \nonumber
\end{eqnarray}
Thus, the corrections from thermal fermionic excitations are  
\begin{eqnarray}
r \rightarrow r -g_2 \frac{\bar{n}_f}{T}, \quad \lambda \rightarrow \lambda -g_4 \frac{\bar{n}_f}{T}
\end{eqnarray}
There are no additional singular channels from thermal fermionic excitations. For large enough $g_4$, the self-interacting term becomes negative signaling a first order transition.  
Thus, the phase transition is either the dual Ising class one or a first order transition. 
 
We also remark an extension of the the semi-quantum theory to the quantum mechanical one as  
\begin{eqnarray}
\mathcal{S}_{eff} &=&  \int d^3 x d\tau \Big[ \frac{1}{2}(\partial_{\tau} \zeta)^2 + \frac{1}{2}(\nabla \zeta)^2 + \frac{r}{2} (\zeta)^2+\frac{\lambda}{4!} \zeta^4 \nonumber\\ 
&+&  f_{\alpha}^{\dagger}(\partial_{\tau}+ \epsilon_f(-i\nabla)) f_{\alpha} + g_2 \,\zeta^2 f_{\alpha}^{\dagger} f_{\alpha}+ g_4 \,\zeta^4 f_{\alpha}^{\dagger} f_{\alpha}\Big] \nonumber 
\end{eqnarray}
The boson field $\zeta(x,\tau)$ describes the long wave length fluctuations of the dual Ising field. The linear time derivative term $\zeta \partial_{\tau} \zeta$ vanishes, so the bare dynamical critical exponent is $z=1$.  
Note that the lowest order coupling is $\zeta^2 f_{\alpha}^{\dagger} f_{\alpha}$ , and thus the fermion fluctuations do not modify the dynamical critical exponent.

\begin{figure}
\includegraphics[width=1.7in]{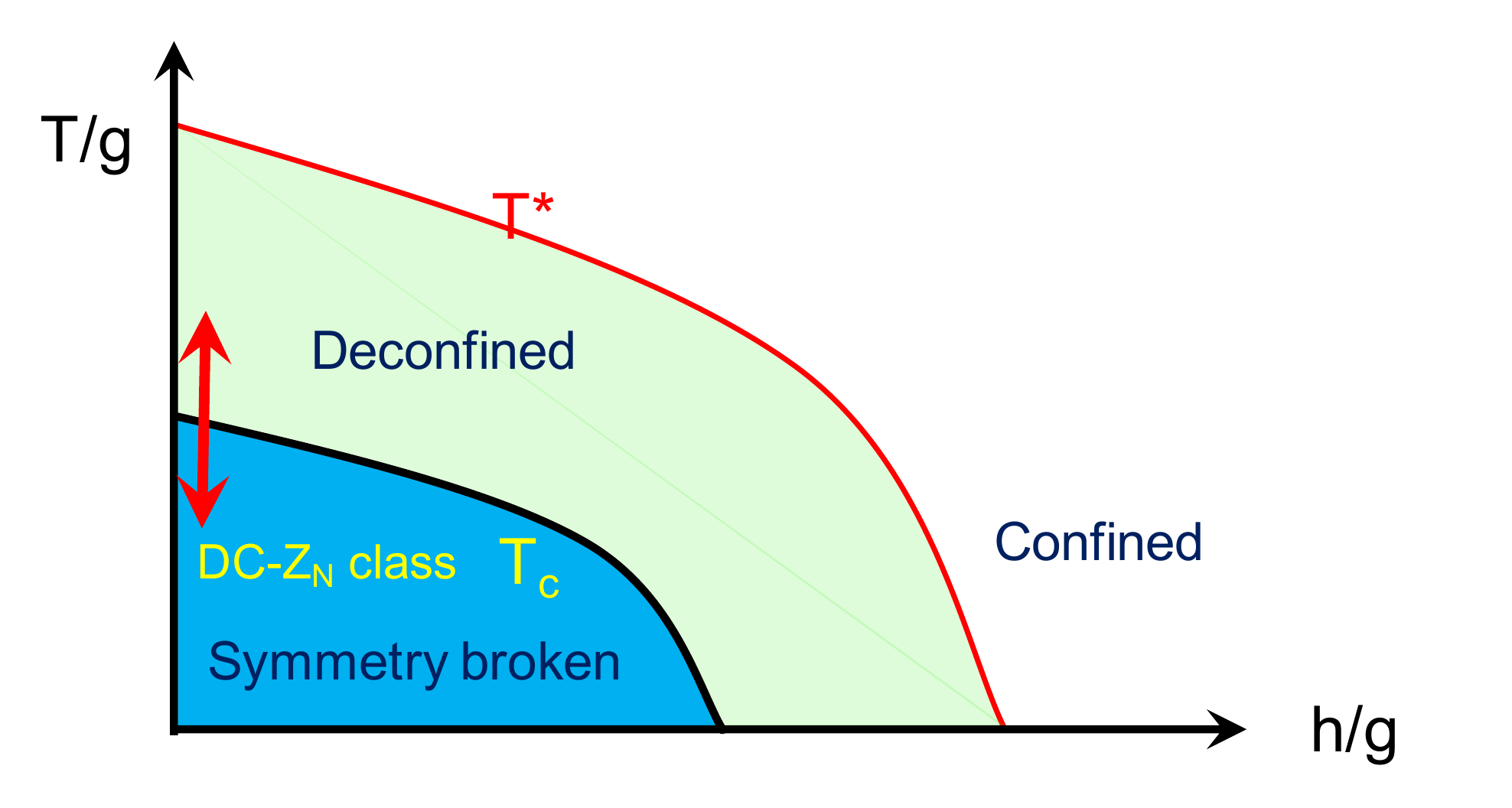}
\includegraphics[width=1.6in]{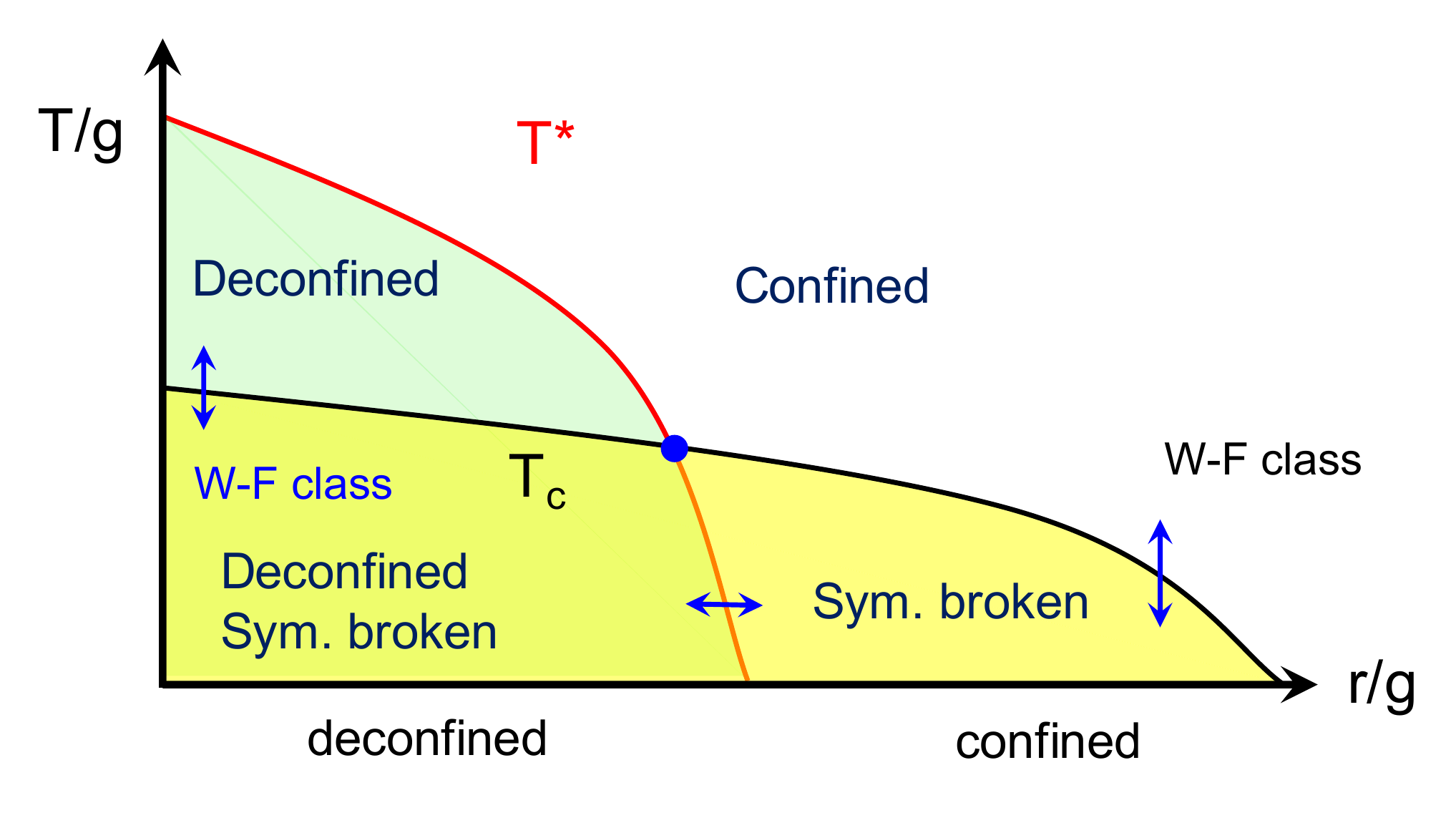}
\caption{Scenarios of symmetry breaking transitions associated with deconfined phases. (Left) The phase transition between a {\it conventional} symmetry broken phase to a deconfined phase is illustrated. (Right) Symmetry breaking transitions and deconfinement transitions are decoupled. In a symmetry broken phase, deconfined excitations may (right) or may not (left) exist.  } \label{visons}
 \end{figure}

\section{Scenarios of symmetry breaking transitions associated with deconfined phases and extension to larger gauge groups}
In the main-text, we discuss the unconventional universality classes (DC-$Z_N$ or DC-U(1)) of symmetry breaking transitions at non-zero temperatures. 
There are other possibilities to realize thermal phase transitions and we introduce them in this section. 

First, the transition connects a symmetric broken {\it conventional} phase to a deconfined phase both thermally and quantum-mechanically. 
There is no topological structure in a symmetry broken phase, and the deconfined phase contain the symmetry broken phase. Only $4\pi$ vortex proliferation is important.  
Second, the conventional Wilson-Fisher universality class may be possible if a symmetry broken phase contains deconfined excitations. The order parameter is gauge-neutral, and its condensation does not alter the gauge sector. All the symmetry breaking transitions are in the Wilson-Fisher universality class. 

It is straightforward to generalize our discussion for $Z_M$ gauge groups. 
The model Hamiltonian with global $Z_N$ symmetry and $Z_M$ gauge symmetry is
\begin{eqnarray}
H= -t \sum_{\langle i,j \rangle} (\mu_{ij} e^{i \frac{\theta_i - \theta_j}{M}} +h.c. )- u \sum_i \cos( N \theta_i) + g V_{F}(\{ \mu_{ij}\}), \nonumber
\end{eqnarray}
where a $Z_M$ gauge field $\mu_{ij}$ is introduced. $V_F$ describes a gauge flux energy term. In $3d$, the confinement-deconfinement transition exists, so we can focus on the deconfined phase by taking the limit $g \gg T, t, u$. 
Using the similar analysis of the $Z_2$ lattice gauge structure, the order parameter becomes higher order powers ($e^{i \theta_i} = (e^{i \theta_i/M})^M$), which give the critical exponents, $\beta_M$,
\begin{eqnarray}
\beta_3 =1.42,  \quad \beta_4=2.09,
\end{eqnarray}
\cite{Vicari} and thus the order parameters may show super-linear onsets below critical temperatures. 
Note that the $Z_2$ gauge structure critical exponent is $\beta_{2} = 0.83$.
Other critical exponents are easily determined by using the scaling relations with $\alpha_M=-0.015$ as shown in Table \ref{T2}. 
The larger gauge structure is associated with the suppression of higher order vortices, and more detailed analysis will be presented in future works.

\begin{table}[tb!]
\begin{tabular} {|c|c|c|c|c|c|c|}
\hline \hline
DC-U(1) in $3d$ & $\alpha$  & ~~$\beta$~~  & ~~$\gamma$~~ & ~~$\nu$~~ & ~~$\eta$~~ & ~~$\delta$~~  \\ \hline \hline
$Z_2$ gauge   & $-0.015$    & $0.83$ & $0.35$ & $0.67$ & $1.47$& $1.43$
 \\ \hline 
$Z_3$ gauge  & $-0.015$    & $1.42$ & $-0.83$ & $0.67$ & $3.23$& $0.42$
 \\ \hline 
$Z_4$ gauge   & $-0.015$    & $2.09$ & $-2.17$ & $0.67$ & $5.22$& $-0.04$  \\ \hline \hline 
\end{tabular}
\caption{Universality classes in $3d$ with different gauge groups. The $Z_N$ potentials are irrelevant in every class. The notations are the same as in the main-text. 
}
\end{table} \label{T2}

\begin{figure}
\includegraphics[width=2.5in]{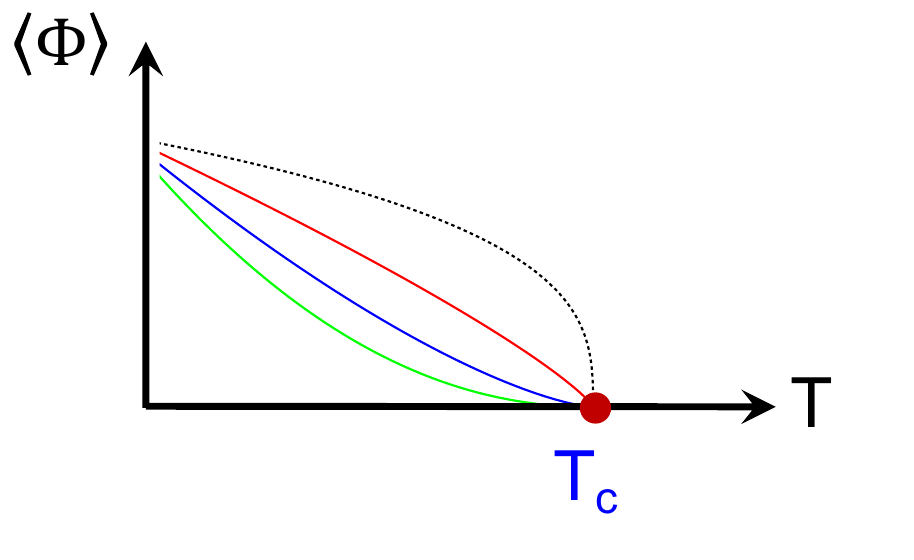}
 \caption{ Different onsets of $Z_2$ order parameters associated with different gauge structures. The dotted black line is for the Ising universality class under the Landau paradigm with $\beta_{Ising}=0.33$. The red, blue, green lines are for DC-$Z_2$ with $Z_2$, $Z_3$, $Z_4$ gauge structures whose critical exponents are $\beta_2=0.83$, $\beta_3=1.42$, and $\beta_4=2.09$, respectively. }  
 \end{figure}

\section{Implications to the hidden order phase in URu$_2$Si$_2$}
Our deconfined thermal transitions in metals do not require any broken symmetries though specific heat experiments show singular temperature dependences such as  continuous dual Ising or discontinuous transitions. 
It is tempting to apply our theories to mysterious hidden order transitions of URu$_2$Si$_2$, which has been investigated by a number of the proposed theories. In contrast to the previous theories, our transitions are {\it intrinsically} independent of symmetries, and therefore it is impossible to measure with experimental probes of broken symmetries.   
Note that some recent experiments, on the other hand, report rotational symmetry breaking from the tetragonal symmetry down to the orthorhombic one in URu$_2$Si$_2$ at the hidden order temperature. 

We propose the presence of the two transitions, a symmetric deconfined transition at the hidden order temperature and the rotational symmetry breaking transition at lower temperature, to explain the rotational symmetry breaking. 
The pattern of the rotational symmetry breaking is in the $Z_2$ class from the tetragonal symmetry to the orthorhombic one, so we can use the enlarged universality class with $\alpha <0$ and $\beta_M$ (see the previous section). The negative value of $\alpha$ indicates that it may be difficult for specific heat experiments to distinguish the two transitions. We stress that the value of $\beta_M$ is much bigger than any other conventional ones in $3d$ such as one of the Ising class ($\beta_{Ising}=0.33$) and seems to fit the magnetic torque data better. 
Further works on more quantitative analysis and comparison with experiments are highly desired.


 \end{document}